\documentclass[sigconf,natbib=false,anonymous=false]{acmart}

\renewcommand\footnotetextcopyrightpermission[1]{}
\setcopyright{none}
\acmConference[arXiv Preprint]{}{June 2025}{}
\acmYear{}
\acmISBN{}
\acmDOI{}

\usepackage{tikz}
\usepackage{pgfplots}
\pgfplotsset{compat=1.18}
\usetikzlibrary{calc,decorations.pathreplacing}
\usepackage{courier}
\usepackage[ruled,vlined]{algorithm2e} 
\usepackage{adjustbox}
\usepackage{multirow}
\usepackage{xcolor}
\usepackage{subcaption}
\usepackage{algpseudocode}
\usepackage{setspace}

\RequirePackage[
  datamodel=acmdatamodel,
  style=acmnumeric,
]{biblatex}
\begin{document}

\title{SEP-GCN: Leveraging Similar Edge Pairs with Temporal and Spatial Contexts for Location-Based Recommender Systems}

\author{Tan Loc Nguyen}
\orcid{0009-0002-4926-398X}
\affiliation{%
  \institution{Faculty of Information Technology, Ton Duc Thang University}
  \city{Ho Chi Minh City}
  \country{Vietnam}
}
\email{232805002@student.tdtu.edu.vn}

\author{Tin T. Tran}
\orcid{0000-0003-4252-6898}
\authornote{Tin T. Tran is the corresponding author.}
\affiliation{%
  \institution{Artificial Intelligence Laboratory, Faculty of Information Technology, Ton Duc Thang University}
  \city{Ho Chi Minh City}
  \country{Vietnam}
}
\email{trantrungtin@tdtu.edu.vn}

\renewcommand{\shortauthors}{Nguyen Tan Loc and Tin T. Tran}

\begin{abstract}
Recommender systems play a crucial role in enabling personalized content delivery amidst the challenges of information overload and human mobility. Although conventional methods often rely on interaction matrices or graph-based retrieval, recent approaches have sought to exploit contextual signals such as time and location. However, most existing models focus on node-level representation or isolated edge attributes, underutilizing the relational structure between interactions. We propose SEP-GCN, a novel graph-based recommendation framework that learns from pairs of contextually similar interaction edges, each representing a user-item check-in event. By identifying edge pairs that occur within similar temporal windows or geographic proximity, SEP-GCN augments the user-item graph with contextual similarity links. These links bridge distant but semantically related interactions, enabling improved long-range information propagation. The enriched graph is processed via an edge-aware convolutional mechanism that integrates contextual similarity into the message-passing process. This allows SEP-GCN to model user preferences more accurately and robustly, especially in sparse or dynamic environments. Experiments on benchmark data sets show that SEP-GCN consistently outperforms strong baselines in both predictive accuracy and robustness.
\end{abstract}



\maketitle

\begin{center}
\textbf{This is a preprint submitted to arXiv.}
\end{center}

\section{INTRODUCTION}

Recommender systems have become essential to modern digital platforms, driving personalized experiences in e-commerce, social networks, and content streaming. Traditional methods such as collaborative filtering and content-based approaches provide foundational frameworks, yet struggle to capture complex high-order relationships. Graph-based methods \cite{hamilton2017inductive,kipf2017semi} address this by modeling user-item interactions as graphs, unlocking richer structural patterns. Graph Neural Networks (GNNs), especially Graph Convolutional Networks (GCNs), have shown strong performance in recommendations by allowing nodes to aggregate neighbor information. Models like NGCF \cite{he2017neural} and LightGCN \cite{he2020lightgcn} leverage high-order connectivity, but often under-use rich edge-level context such as time and location. These spatio-temporal signals are critical to understanding dynamic user behavior.

Contextual edge information—such as transaction time and location—can significantly enhance recommendation quality. Frequent interactions during certain times or within specific regions imply strong behavioral patterns. While models like KGAT \cite{wang2019multi} and TAGNN \cite{wu2021comprehensive} explore such contexts, they often lack explicit propagation of edge-level signals across the graph. A further challenge lies in capturing long-range dependencies. Real-world graphs often contain densely connected communities with sparse inter-cluster links, limiting GNNs to a few hops due to oversmoothing and computational costs \cite{zhang2018link}. As a result, meaningful signals from distant but relevant nodes may be lost.

To overcome these limitations, we propose \textbf{Similar Edge Pair learning with a bare-bones GCN framework (SEP-GCN)}. Instead of focusing solely on nodes or individual edges, SEP-GCN learns from \textit{similar pairs of interaction edges} that share spatio-temporal context. We construct a secondary graph where edges are connected based on contextual similarity, allowing richer relational learning.

This edge-pair-centric approach brings two key benefits: (1) it explicitly leverages spatio-temporal edge context more effectively than prior GNNs, and (2) it enables efficient propagation across topologically distant yet semantically similar regions. SEP-GCN thus captures complex behavioral patterns while avoiding oversmoothing and scaling issues, achieving strong gains over state-of-the-art baselines in contextual recommendation tasks.

\section{RELATED WORK}
\subsection{Graph Convolutional Networks in Recommender Systems}
Graph Neural Networks (GNNs) have gained significant attention in the field of recommendation systems (RS) due to their ability to model complex relationships between entities. Traditional recommendation methods, such as collaborative filtering and matrix factorization, rely on learning latent representations from user-item interaction matrices. However, these methods often fail to capture the higher-order connectivity and graph-structured data inherent in many recommendation scenarios. GNN-based approaches address these limitations by leveraging the power of graph embeddings to propagate information over user-item interaction graphs.

The introduction of GCNs into RS, as demonstrated in NGCF (Neural Graph Collaborative Filtering) \cite{wang2019ngcf}, laid the groundwork for encoding user-item interactions through iterative message passing mechanisms. NGCF improves representation learning by incorporating first-order and higher-order neighborhood information. Similarly, LightGCN \cite{he2020lightgcn} proposed a simplified GCN framework, eliminating redundant operations such as non-linear activations and feature transformations, resulting in improved computational efficiency and performance. Beyond basic GCNs, Graph Attention Networks (GATs) \cite{velickovic2018graph} have been used in RS to adaptively assign importance weights to graph edges, highlighting significant user-item interactions. These advances demonstrate that GNNs effectively capture the rich structural and contextual information in recommendation tasks.

Furthermore, heterogeneous graph neural networks (HGNNs) have been explored for complex RS scenarios where user-item interactions are augmented with side information, such as user demographics or item attributes. Methods like Heterogeneous Graph Collaborative Filtering (HGCF) extend the GNN framework to incorporate multiple types of nodes and edges, improving the precision of the recommendation in multifaceted datasets \cite{zhang2020deep}. 

\subsection{Temporal and Partial Recommender Systems}
Location-Based Recommender Systems (LBRS) leverage spatio-temporal data to provide personalized recommendations, addressing challenges such as data sparsity, dynamic preferences, and spatial-temporal interplay. The temporal aspect is crucial, as user behaviors differ by time, e.g., morning coffee shops vs. evening restaurants \cite{Xu2020, Ye2011}. To model this, researchers often segment the week into discrete slots, such as the 168 weekly slots proposed in \cite{Xu2020}, simplifying computations while retaining behavioral trends. Temporal granularity helps to identify repeat patterns \cite{Ye2011} and enhances predictive performance in models such as collaborative filtering \cite{Koren2009} and spatial-temporal classification \cite{Li03032024}. However, these approaches treat temporal features as static inputs and lack explicit mechanisms to propagate temporal signals across the graph structure.

Sequential LBRS analyze user mobility over time. Although traditional methods such as Markov chains are limited by scalability and shallow historical modeling, deep learning models such as RNN and LSTM \cite{Hidasi2015} have improved sequential modeling but struggle with generalization in dynamic or sparse contexts. Session-based LBRS further leverages GNNs, such as SR-GNN \cite{Wu2019}, to capture short-term item transitions. For the next-POI recommendation, spatial-temporal GNNs such as STGN \cite{STGN2020}, DeepMove \cite{DeepMove2018} and STAN \cite{10.1145/3442381.3449998} model user trajectories using attention mechanisms. Although effective, these models focus primarily on modeling individual user sequences and do not exploit the structural similarity between different user-item interactions in shared contexts.

Advanced LBRS adopt dynamic graph structures to capture evolving interactions. Temporal Graph Neural Networks (TGNNs), such as Jodie \cite{Kumar2019} and TGAT \cite{Xu2020}, dynamically update the embeddings of users and items over time. Context-aware GNNs further enrich the recommendation by incorporating external factors such as weather and traffic. For example, Tran et al. \cite{tiis:100957} improved location recommendation by integrating social ties and spatial proximity. Despite these advances, most models still rely on node-level or individual edge representations, overlooking the rich relational patterns between similar interactions.

Our work differs from the above by learning from contextually similar interaction pairs rather than isolated nodes or single interactions. By explicitly modeling temporal and spatial similarity between user-item edges, our method enables more expressive representation learning and long-range dependency propagation, which are often underexplored in prior LBRS frameworks.

\subsection{Edge-enhanced Graph Neural Networks}
Edge-enhanced GNNs extend traditional graph neural architectures by incorporating edge features, enabling richer relational modeling beyond node-only representations. These models have shown improvements in tasks such as link prediction and graph classification using edge-specific information. For example, Lin et al. \cite{Lin2022ImprovingGC} proposed an edge-aware GNN for collaborative filtering, where edge attributes helped capture sparse user-item interactions. Wang et al. \cite{Wang2023EEDN} introduced the Enhanced Encoder-Decoder Network (EEDN), combining hypergraph convolutions with edge features to address cold-start problems in point-of-interest recommendations.

Recent studies further co-embed nodes and edges. Zhou et al. \cite{Zhou2023} explored edge-node co-embedding for relational link prediction, while Gong and Cheng \cite{Gong2019EdgeFeatures} demonstrated that explicitly modeling edge features improves node classification. Chen and Chen \cite{Chen2021EGAT} extended the attention mechanism to edge attributes via Edge-Featured Graph Attention Networks (EGAT), and Cai et al. \cite{Cai2021EGNAS} proposed EGNAS to automate the discovery of edge-aware architectures.

Frameworks such as PyTorch Geometric (PyG) \cite{PyG2023} now support edge attributes in the core GNN layers like \texttt{MessagePassing}, facilitating the development of edge-aware models in various domains. Despite these advances, most approaches treat edge features independently and do not model structural similarity or contextual alignment between edges, particularly in spatio-temporal settings.


\section{PRELIMINARIES}
\subsection{Definitions}

\textbf{Definition 1} (Check-in matrix). Given a set of recorded users $\mathcal{U}=\{u\}$ and set of items $\mathcal{I}=\{i\}$, the check-ins matrix $\textbf{R} \in \{0,1\}^{\vert \mathcal{U} \vert x \vert \mathcal{I} \vert}$, where each element $\textbf{R}_{i,j}=1$ denotes a check-in by user $u_i$ in place $i_j$, otherwise $\textbf{R}_{i,j}=0$. It should be noted that items in location based recommender systems are general term for public places such as museums, train stations, restaurants, parks, and libraries.\\

\noindent
\textbf{Definition 2} (Check-in graph). We define a bipartite interaction graph as $\mathcal{G} = (\mathcal{V}, \mathcal{E}, \mathcal{X})$, where $\mathcal{V}=\{\mathcal{U} \bigcup \mathcal{I}\}$ denotes the set of vertices, $\mathcal{E} = \{e_{u,i} | u \in \mathcal{U}, i \in \mathcal{I}, \textbf{R}_{i,j}=1\}$ denotes the set of edges, and $\mathcal{X} \in \mathbb{R}^{\vert \mathcal{E} \vert x \vert \mathcal{E} \vert}$ represents the matrix capturing the correlations between the edges in $\mathcal{E}$.\\

\noindent
\textbf{Task 1} (Top-k recommendation). The task of generating a ranked list of the top-k items that are most likely to be relevant to a user based on their preferences, historical interactions, or contextual information. This approach is widely applied in recommendation systems to provide users with a concise selection of suggestions, ensuring both relevance and diversity within the limited list of k items.


\subsection{Similarity Edge Pairs}

\begin{figure}[ht]
\begin{adjustbox}{width=\columnwidth} 
	\begin{tikzpicture}[scale=0.8, every node/.style={scale=0.8}]

			\draw[->] (0,0) -- (12.5,0) node[above] {Time (hours)};
			\draw[->] (0,0) -- (0,4) node[left] {Check-in};
			
			\draw (0,0) -- (0,-0.2) node[below] {0};
			\draw (1.125,0) -- (1.125,-0.2) node[below] {6};
			\draw (2.25,0) -- (2.25,-0.2) node[below] {12};
			\draw (4.5,0) -- (4.5,-0.2) node[below] {24};
			\draw (9,0) -- (9,-0.2) node[below] {48};
			\draw (11.8,0) -- (11.8,-0.2) node[below left=-2pt] {168};
			\node at (6.75,-0.5) {\dots};
			\node at (10.4,-0.5) {\dots};
			
			\draw[decorate,decoration={brace,mirror,amplitude=6pt},blue] 
			(0,-0.8) -- (4.5,-0.8) node[midway,below=10pt,text=blue] {Monday};
			\draw[decorate,decoration={brace,mirror,amplitude=6pt},red] 
			(4.5,-0.8) -- (9,-0.8) node[midway,below=10pt,text=red] {Tuesday};
			
			\foreach \x in {0.1, 1.3,  3.7, 5.5, 6.8, 8.4} 
			\node[draw,fill,inner sep=1pt,circle] at (\x,1) {}; 
			\foreach \x in {1,2,11} 
			\node[draw,fill=red,inner sep=2pt,circle] at (\x,1) {};

			\foreach \x in {1,2,11} 
			\node[draw,fill=red,inner sep=2pt,circle] at (\x,2) {};
			\foreach \x in {0.5,4,9} 
			\node[draw,fill=green,inner sep=2pt,circle] at (\x,2) {};
			\foreach \x in {0.3, 1.5, 4.2, 5.1, 6.4, 7.9, 9.6, 11.2} 
			\node[draw,fill,inner sep=1pt,circle] at (\x,2) {};
			
			\foreach \x in {0.5,4,9} 
			\node[draw,fill=green,inner sep=2pt,circle] at (\x,3) {};
			\foreach \x in {0.4, 1.7, 3.5, 5.3, 6.8, 8.1, 9.2} 
			\node[draw,fill,inner sep=1pt,circle] at (\x,3) {};
			
			\node[anchor=east] at (-0.5,1) {Edge 1};
			\node[anchor=east] at (-0.5,2) {Edge 2};
			\node[anchor=east] at (-0.5,3) {Edge 3};
 
	\end{tikzpicture}
    		\end{adjustbox}  
	\Description[The chart displays check-in patterns across three edges over a week, with time on the x-axis (in hours) and edge labels on the y-axis. Green and red circles indicate specific events or types of check-ins, while black dots represent regular activity.]{The chart shows the temporal distribution of check-ins across three edges (Edge 1, Edge 2, and Edge 3) over a week. Time is measured in hours (x-axis), and edges are labeled along the y-axis. Each edge has specific check-ins, shown as dots, with color coding to differentiate event types (e.g., red and green). The distribution highlights daily activity patterns or scheduling regularities.}
	\caption{Weekly Time Slots and Time Slot Sets of Check-ins.}
	\label{fig:time_slots}

\end{figure}
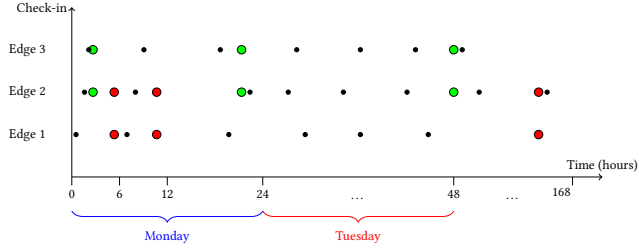

\noindent
\textbf{Definition 3} (Time slot set of a check-in).
For each edge (check-in) $e_{u,i}$ represents a check-in at location $i$ at specific timestamps in a list $t$, where each element in $t$ is an integer belonging to the range $[0, 167]$, where each week is divided into 7 days and each day is further divided into 24 hours. This partitioning scheme uses discrete \textit{slots} to represent units of time within a week, each slot indicating a specific time point (from 0 to 167). Each check-in $e_{u,i}$ corresponds to a specific \textit{time slot}, capturing the exact moment of check-in at the location $i$. The set of all time slots for a location $i$ forms its \textit{time slot set}, which represents cumulative check-ins at different time points. This approach is commonly used in studies involving spatial-temporal data in recommender systems \cite{10.1145/3442381.3449947,10.1609/aaai.v38i8.28697,YEGANEGI2024127104}.\\

We illustrate the time slot set of 3 check-ins in Figure \ref{fig:time_slots}. The time slots set of Edge 1 and Edge 3 are disjoint, since there are no overlapping time slots between them. This indicates that the two locations corresponding to these edges have no similarity. For instance, if Edge 1 represents a coffee shop and Edge 3 represents a library, the disjoint time slots suggest that customers visiting the coffee shop do not visit the library during the same time intervals, indicating distinct user behaviors or purposes for the locations. Meanwhile, the pair of Edge 1 and Edge 2, or Edge 2 and Edge 3, overlap in some time slots, indicating a correlation in their time intervals. For example, if Edge 1 represents a coffee shop and Edge 2 represents a bakery, the overlapping time slots suggest that customers might visit both locations during breakfast hours.\\

\noindent
\textbf{Definition 4} (Geographical Distance Between Locations). Consider two locations on the Earth's surface represented by their geographical coordinates $(X_1, Y_1)$ and $(X_2, Y_2)$, where $X_1, X_2$ denote the latitudes and $Y_1, Y_2$ denote the longitudes of the respective locations in degrees. The geographical distance between these two points is defined using the haversine formula, which accounts for the spherical shape of the Earth, as shown in Equation (\ref{eq:haversin}). This formula provides the great-circle distance, which is the shortest path between two points on the surface of a sphere.

\begin{equation} \label{eq:haversin}
	d=2r\sin^{-1}\left(\sqrt{\sin^2\left(\frac{X_2-X_1}{2}\right)+\cos(X_1)\cos(X_2)\sin^2\left(\frac{Y_2-Y_1}{2}\right)}\right)
\end{equation}

\noindent
\textbf{Definition 5} (Similarity Edge Pairs). 

\begin{figure}[ht]
	\includegraphics[width=\linewidth]{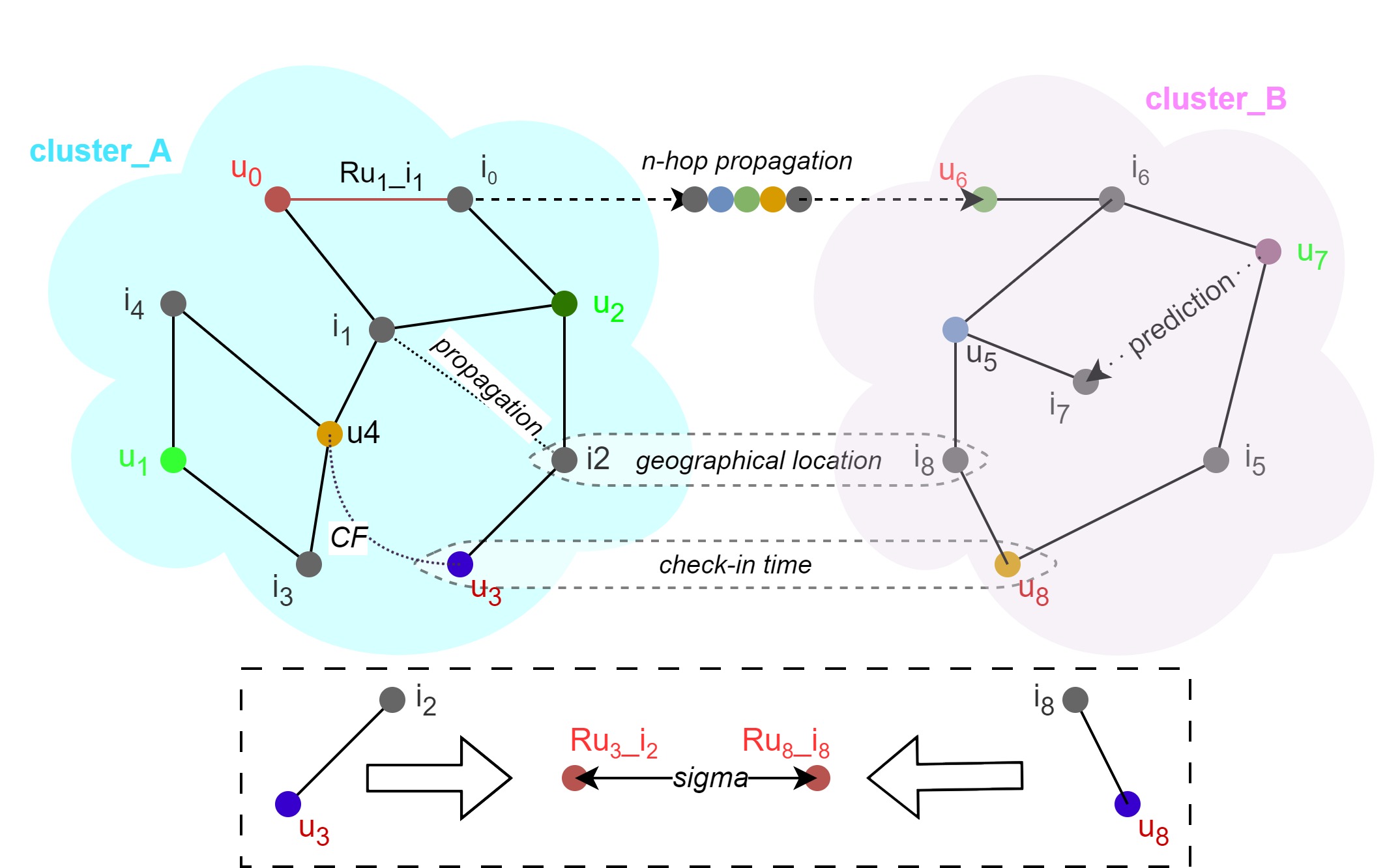}
	\caption{Edge $(u_3, i_2)$ and $(u_8, i_8)$ is a potential SEP.}
	\label{fig:SEP}
	\Description[The figure illustrates the SEP (Spatial-Temporal Edge Propagation) mechanism, highlighting two clusters of users and items (clusterA and clusterB), where edges (u3, i2) and (u8, i8) are identified as potential SEPs based on geographical distance, check-in time, and n-hop propagation.]{Figure 2 demonstrates the graph structure integrating user and item clusters, illustrating spatial-temporal relationships and signal propagation between nodes. In cluster_A, user u3 interacts with item i2 through connections defined by check-in time and geographical proximity. Similarly, cluster_B showcases the interaction between u8 and i8. The edges (u3, i2) and (u8, i8) are identified as potential SEPs due to their strong relational signals, represented by propagation scores Ru3,i2 and Ru8,i8. Dashed lines represent the n-hop propagation mechanism, supported by additional contextual factors such as geographical location and temporal associations. The lower section of the figure provides a detailed depiction of how sigma weights are calculated for the edges, emphasizing the role of SEP in improving interaction predictions.}
\end{figure}

Given two edges of interaction $e_1 = (u_1, i_1)$ and $e_2 = (u_2, i_2)$ in the set of edges $\mathcal{E}$, we define a \textit{Similar Edge Pair} (SEP) with similarity weight $\sigma(e_1, e_2)$ if and only if the following two conditions are satisfied:

\begin{enumerate}
	\item The time slot sets of $e_1$ and $e_2$, denoted as $t_1$ and $t_2$, satisfy $t_1 \cap t_2 \neq \emptyset$. We compute these slots based on the users’ \textit{local time zones} rather than using a global timestamp. This adjustment improves the contextual consistency of temporal patterns and effectively reduces the inclusion of geographically irrelevant locations that would otherwise appear simultaneously in a global time window.
	
	\item The geographical distance between $l_1$ and $l_2$, the locations associated with $e_1$ and $e_2$, is evaluated using a soft exponential decay function. Specifically, we first compute the median distance of all visited locations for each user. Then, for any pair of edges, we define the similarity score $\sigma$ between $e_1$ and $e_2$ by Equation (\ref{eq:cal_sigma}), which integrates both temporal overlap and spatial proximity.
	This function ensures that:
	\begin{itemize}
	    \item Location pairs with a distance equal to the user's median receive a moderate similarity score $\alpha$.
	    \item The similarity increases toward $1.0$ as the distance decreases below the median. In contrast, similarity decays exponentially towards $0.0$ as the distance increases beyond the median.
	\end{itemize}
\end{enumerate}

\begin{equation} \label{eq:cal_sigma}
	\sigma(e_1, e_2)=\exp\left(\frac{d(i_1, i_2)}{median(\forall d_i\in \mathbb{D})}*\ln\alpha\right)
\end{equation}
where $d(i_1, i_2)$ is the geographic distance between the two locations, calculated using Equation (\ref{eq:haversin}), $\alpha$ is the correlation value at the median point, and $\mathbb{D}$ represents the set of all distance of all edges in $\mathcal{E}$.

Figure~\ref{fig:SEP} illustrates this concept: edges $(u_3, i_2)$ and $(u_8, i_8)$ form a valid SEP due to both overlapped time slots and spatial proximity. In contrast, other edges lacking either condition are excluded. This mechanism enables SEP-GCN to selectively propagate contextual signals between interaction pairs, even when they are topologically distant in the original graph.

This mechanism acts as a personalized radius filter, effectively suppressing spatially outlying interactions while maintaining a long-tailed distribution of location pairs based on distance. It emphasizes closer interactions without requiring a fixed threshold.\\

\noindent
\textbf{Definition 6} (SEP matrix):
The matrix $\mathcal{X} \in \mathbb{R}^{\vert \mathcal{E} \vert \times \vert \mathcal{E} \vert}$ represents the correlation between edge pairs in the set $\mathcal{E}$, where each element $x_{i,j}$ is computed using the similarity function $\sigma$ in Equation \eqref{eq:cal_sigma}, based on the edges $e_i$ and $e_j$. The matrix $\mathcal{X}$ is exhibit symmetry, reducing storage and computational costs. In practice, $\mathcal{X}$ is often a sparse matrix, as not all edge pairs share significant relationships. To optimize computation, the construction of matrix $\mathcal{X}$ is described in Algorithm 1. In that process, only relevant edge pairs based on temporal or spatial constraints is focused. This ensures that the matrix is efficiently and accurately constructed, laying a solid foundation for further analysis in the model.

\begin{algorithm}
\caption{Compute Matrix $\mathcal{X}$ Containing Correlation Values Between Potential Edge Pairs}
\label{alg:compute_correlation_matrix}
\SetAlgoLined
\KwIn{$E$: Edge set, $T$: Time set, $L$: Location set}
\KwOut{Matrix $\mathcal{X}$ with correlation values}

Initialize $\mathcal{X} \gets$ empty matrix\;

\ForEach{edge $i \in E$}{
    Get $time_i \in T$\;
    Get $loc_i \in L$\;
}

\ForEach{pair of edges $i, j \in E \times E$}{
    \If{$time_i$ and $time_j$ have non-empty intersection}{
        $\sigma \gets \text{normalized\_distance}(loc_i, loc_j)$\;
        $\mathcal{X}.\text{row}.\text{append}(j)$\;
        $\mathcal{X}.\text{col}.\text{append}(i)$\;
        $\mathcal{X}.\text{data}.\text{append}(\sigma)$\;
    }
}

$\mathcal{X} \gets F.\text{normalize}(\mathcal{X})$\;
\Return $\mathcal{X}$\;
\end{algorithm}

			                
\section{METHODOLOGY}
GCN-based models remain pioneering models in the field of recommender systems. We represent the features of the graph nodes, including the user nodes and the item nodes, as the feature embeddings $E_U$ and $E_I$. The SEP information should be incorporated immediately after each propagation step. Therefore, we also encode SEP into embeddings $E_{SEP}$ and aggregate the signals using a function whose parameters can be adjusted. When updating the user embeddings $E_U$, all the items associated with the potential edges generate direct reinforcement signals. Similarly, when updating the embeddings of the items $E_I$, users also contribute potential SEP signals to the items, even though they have not interacted directly.

\begin{figure*}
\includegraphics[width=\textwidth,keepaspectratio]{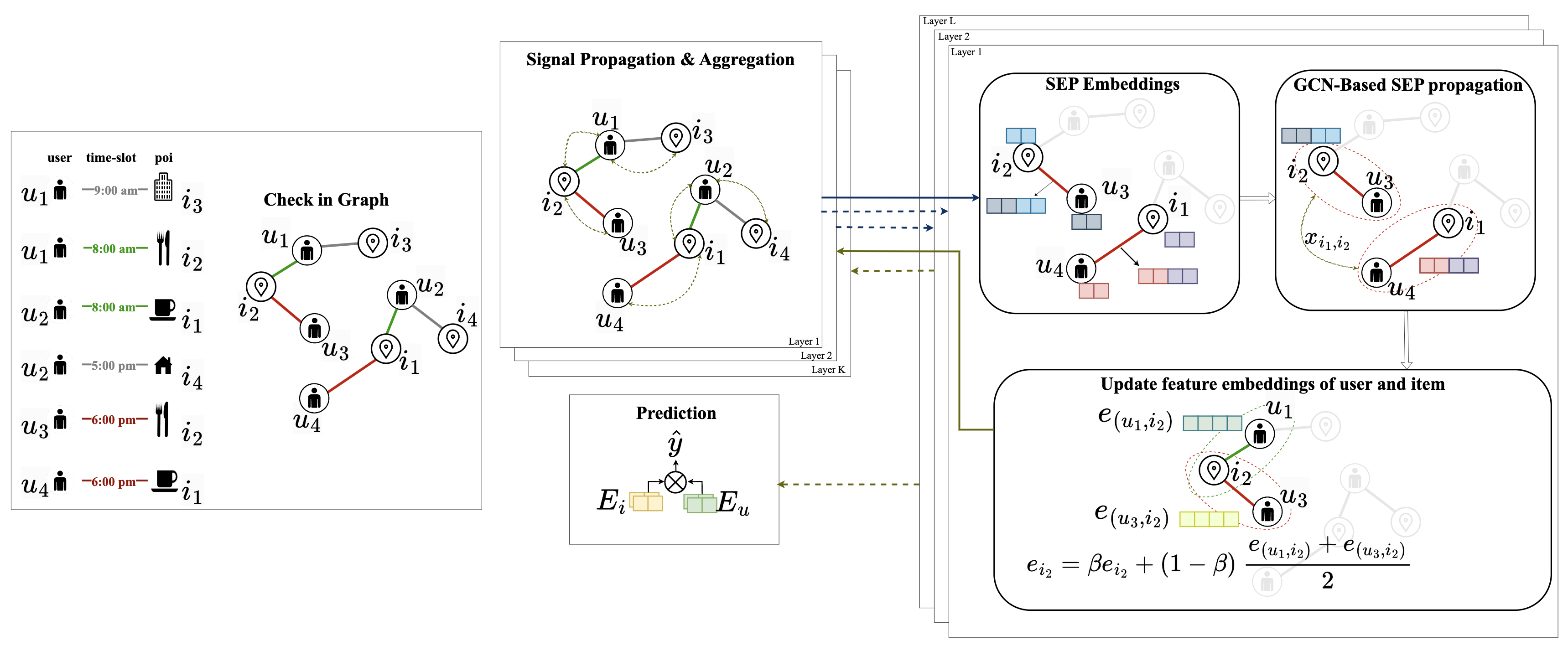}
\caption{The proposed SEP-GCN model architecture, incorporating spatiotemporal edge similarity into GCN-based propagation}
\label{fig:overview}
\Description[The figure illustrates the SEP-GCN framework, highlighting the check-in graph construction, signal propagation, embedding updates, and final prediction process.]{The figure provides a comprehensive overview of the SEP-GCN framework. On the left, the check-in graph is constructed, incorporating users, items (points of interest), and temporal interactions with distinct time slots. Users and items are connected through edges annotated with temporal and spatial context. The central section demonstrates the signal propagation and aggregation mechanism across multiple GCN layers, where user-item relationships are enhanced with SEP embeddings and GCN-based propagation. On the right, the embedding update module integrates contextual edge information into the user and item feature embeddings through a weighted combination formula. Finally, the enriched embeddings are fed into the prediction module to generate personalized recommendations.}
\end{figure*}

The overall architecture of SEP-GCN is illustrated in Figure \ref{fig:overview}. The framework consists of two main modules: (1) a standard LightGCN-based user-item propagation layer (left), where embeddings are updated through neighbor aggregation in the user-item bipartite graph \cite{he2020lightgcn}; and (2) a novel Similar Edge Pair (SEP) module (right), where edge-level embeddings, constructed via concatenation of user and item embeddings, are propagated through an edge similarity graph. The SEP graph is defined over pairs of interactions with shared temporal and/or spatial context. After each GCN layer, the SEP embeddings are used to refine user and item embeddings through a weighted update mechanism. This dual propagation approach enables SEP-GCN to capture high-order context-aware signals that improve recommendation accuracy.

{\subsection{GCN-based propagation}}
\label{sec:gcn_user_item_propagation}
\subsubsection{Initialization of feature embeddings of Users and Items}
The input to the model consists of user and item features represented in a vector space. Let $e_u$ denote the user features and $e_i$ denote the item features, both randomly initialized from a normal distribution. In matrix form, the complete set of embeddings used during propagation is represented as 
$E^{(0)} \in \mathbb{R}^{(n+m) \times d}$, 
where $E^{(0)}$ contains $n$ user embeddings and $m$ item embeddings, with $d$ representing the embedding size.

\begin{equation}
    E^{(0)}=E_U^{(0)} \parallel E_I^{(0)}=[e_{u_1}^{(0)},\dotso ,e_{u_n}^{(0)},e_{i_1}^{(0)},\dotso,e_{i_m}^{(0)}]
\label{eq:init_E}
\end{equation}

\subsubsection{Signal Propagation and Aggregation}
User and item features are propagated as embeddings $e_u^k$ (representing the user features) and $e_i^k$ (representing the item features) at the $k$-th layer, calculated using Equation (\ref{embed_equations}). Here, $\mathcal{N}_{u}^{A}$ and $\mathcal{N}_{i}^{A}$ denote the neighbors of user $u$ and item $i$ in the adjacency matrix $A$, respectively.

\begin{equation} \label{embed_equations}
\begin{aligned}
\mathrm{e}_{u}^{k} &= \sum_{i\in \mathcal{N}_{u}^{A}} \frac{1}{\sqrt{\left| \mathcal{N}_{u}^{A} \right| \left| \mathcal{N}_{i}^{A} \right|}} \mathrm{e}_{i}^{k-1}
\\
\quad
\mathrm{e}_{i}^{k} &= \sum_{u\in \mathcal{N}_{i}^{A}} \frac{1}{\sqrt{\left| \mathcal{N}_{i}^{A} \right| \left| \mathcal{N}_{u}^{A} \right|}} \mathrm{e}_{u}^{k-1}
\end{aligned}
\end{equation}

With the matrix $R$ as an input, we form an adjacency matrix of the User-Item interaction graph is defined as $A = \begin{bmatrix} 0 & R \\ R^\top & 0 \end{bmatrix}$. The normalized symmetric adjacency matrix $\tilde{A}$ is calculated as $ \tilde{A} = D^{-1/2} A D^{-1/2} $, where $ D $ represents the degree matrix of $A$. This normalized matrix $\tilde{A}$ is used in the propagation process to ensure that user and item features are appropriately scaled based on their connectivity in the graph. In matrix form, the propagation process for the user and item features can be expressed as Equation (\ref{eq:embed_matrix}).
\begin{equation}
    E^{k} = \widetilde{A}E^{(k-1)} 
\label{eq:embed_matrix}
\end{equation}

In most traditional GCN models, these embeddings $ E^{k} $ serve as input for the next propagation step, with the final embeddings calculated using Equation (\ref{predict}). In our proposed SEP-GCN model, the feature embeddings are updated after each propagation step with the precomputed SEP embeddings. The detailed process for computing the SEP embeddings is provided in the following section.

\subsection{Similar Edge Pair learning}\label{sec:sep_module}
\subsubsection{SEP Embeddings}
The relationship between a user and an item in the check-in graph is represented as a unique edge embedding. This embedding is constructed using the embeddings of the user and item nodes that define the edge. These node embeddings, $e_{u}^{k}$ for the user and $e_{i}^{k}$ for the item, are obtained from the GCN User-Item propagation process. These two embeddings are then concatenated to capture the relationship represented by the edge, as presented in Equation (\ref{eq:edge_embedding}), which form a linked pair in the graph. We choose concatenation when initializing the edge embedding from the two nodes it connects to preserve the unique features of the user and the item. 
\begin{equation}
e_{SEP(u,i)}^l = CONCAT (e_{u}^{k} , e_{i}^{k})
\label{eq:edge_embedding}
\end{equation}
The SEP embedding $E_{SEP}^l$ collected all created $e_{SEP(u,i)}^l$ and participates in the propagation process on graph $\mathbb{X}$.
\subsubsection{GCN-Based SEP propagation}
The matrix $\mathcal{X}$ corresponds to the SEP graph, where each vertex in the SEP graph represents an edge in the original graph $\mathcal{E}$, and the weight of each edge in the SEP graph reflects the correlation between pairs of vertices.
 The SEP embedding is represented as $E_{SEP} \in \mathbb{R}^{(|\mathcal{E}|) \times d}$. We perform embedding propagation using LGC for each $e_{(u,i)}^l \in E_{SEP}^l$ by Equation (\ref{edge_embed_propagation}).
\begin{equation} \label{edge_embed_propagation}
        e_{(u,i)}^{l} =\sum_{v\in \mathcal{N}_{(u,i)}^\mathcal{X}}\frac{1}{ \sqrt{| \mathcal{N}_{(u,i)}^\mathcal{X} ||\mathcal{N}_{(v)}^\mathcal{X}| } }e_{v}\\
\end{equation}
where $\mathcal{N}_{(u,i)}^\mathcal{X}$ denotes the set of adjacent nodes of node $(u,i)$, and $e_v$ is the embedding of node $v$.
In matrix form, the propagation can be written as
$E_{SEP}^{l} = \tilde{\mathcal{X}}\cdot E_{SEP}$,
where $\tilde{\mathcal{X}} = D_{\mathcal{X}}^{-1/2} X D_{\mathcal{X}}^{-1/2}$ is the normalized SEP matrix $\mathcal{X}$, and $D_{\mathcal{X}}$ is the degree matrix of $\mathcal{X}$.
\subsubsection{Update feature embeddings of user and item}


The user and item embedding aggregate features from their connected edges to update their own embeddings. When constructing the edge embedding according to Equation (\ref{eq:edge_embedding}). We perform separate update processes for user embeddings and item embeddings to reduce the risk of over-smoothing. The indices in the matrix $\mathcal{X}$ are also appropriately labeled for this purpose.

Equations (\ref{update_user_from_edges}) and (\ref{update_item_from_edges}) detail the process of updating user and item embeddings. 
Each node's update is normalized by dividing it by the total number of edges connected to the node. Furthermore, adjustable weights $ \alpha $ and $ \beta $ are applied to control the influence of the embedding update.


\begin{equation} \label{update_user_from_edges}
      e_u^k = \alpha \cdot e_u^k + (1 - \alpha)\cdot \sum_{i \in \mathcal{N}_u^A} \frac{e_{(u,i)}^{l}}{|\mathcal{N}_u^A|}
\end{equation}

\begin{equation} \label{update_item_from_edges}
      e_i^k = \beta \cdot e_i^k + (1 - \beta) \cdot \sum_{u \in \mathcal{N}_i^A} \frac{e_{(u,i)}^{l}}{|\mathcal{N}_i^A|}
\end{equation}

After updating the user embedding $e_{u}^{k}$ and the item embedding $ e_{i}^{k}$ with the SEP embedding, these embeddings are used as input for the next layer. The process of GCN-based User-Item propagation and Similar Edge Pair learning is then repeated until layer $K$ is reached.
We implement the update process for user and item embeddings in Algorithm \ref{alg:weighted_aggregation}.


\begin{algorithm}
\caption{Update User and Item Embeddings by SEP Information}
\label{alg:weighted_aggregation} 
\SetAlgoLined
\KwIn{
$E \in \mathbb{R}^{(m+n) \times d}$: Initial embedding matrix of users/items\;
$SEP \in \mathbb{R}^{|\mathcal{E}| \times 2d}$: Edge embedding matrix\;
$\text{IndexMap} = \{(u, i) \mid u, i \in [m+n]\}$: List of edges\;
$\gamma \in [0, 1]$: The weight of the influence of $SEP$ on $E$\;
}
\KwOut{Updated embedding matrix $E$}

Initialize $\text{AggregatedE} \gets \mathbf{0} \in \mathbb{R}^{(m+n) \times d}$\;
Initialize $\text{Count} \gets \mathbf{0} \in \mathbb{R}^{(m+n)}$\;

\For{$k \gets 1$ \KwTo $|\mathcal{E}|$}{
  Let $(u, i) \gets \text{IndexMap}[k]$\;
  $\text{UserSegment} \gets SEP[k, :d]$\;
  $\text{ItemSegment} \gets SEP[k, d:2d]$\;
  $\text{AggregatedE}[u] \gets \text{AggregatedE}[u] + \text{UserSegment}$\;
  $\text{AggregatedE}[i] \gets \text{AggregatedE}[i] + \text{ItemSegment}$\;
  $\text{Count}[u] \gets \text{Count}[u] + 1$\;
  $\text{Count}[i] \gets \text{Count}[i] + 1$\;
}

\For{$j \gets 1$ \KwTo $m+n$}{
  \If{$\text{Count}[j] > 0$}{
    $\text{AggregatedE}[j] \gets \text{AggregatedE}[j] / \text{Count}[j]$\;
    $E[j] \gets \gamma \cdot \text{AggregatedE}[j] + (1 - \gamma) \cdot E[j]$\;
  }
}

\Return $E$\;
\end{algorithm}

\subsection{The prediction and model optimization}
After several iterations of signal propagation, the characteristic vectors of users and items are found by Equation (\ref{embed_final}).
\begin{equation} \label{embed_final}
\begin{aligned}
e_{u}^* &=\frac{1}{K+1}\sum_{k=0}^{K}\mathrm{e}_{u}^{k} \quad
\\
\quad
e_{i}^* &=\frac{1}{K+1}\sum_{k=0}^{K}\mathrm{e}_{i}^{k}
\end{aligned}
\end{equation} 
The characteristic vectors for users and items converge after several iterations of propagation, and the prediction score between user $u_i$ and item $i_j$ can be calculated by Equation (\ref{predict}).
\begin{equation} \label{predict}
\widehat{y}_{ui}={e_{u_i}^*}^\top  e^*_{i_j}
\end{equation}
\\
The Bayesian personalized ranking (BPR) method is the optimal choice to implement the loss function because it is the most effective ranking method for datasets with implicit feedback \cite{rendle2012bprbayesianpersonalizedranking}. We used two sets of observations: $\Omega^+_{ui}$ representing the interacted items and $\Omega^-_{ui}$ representing the non-interacted items. The loss function is calculated by Equation (\ref{loss}).
\begin{equation} \label{loss}
    Loss_{bpr} =  \sum_{\Omega^+_{ui}}\sum_{\Omega^-_{uj}} -ln \sigma ( \widehat{y}_{ui} - \widehat{y}_{uj} ) +  \lambda  \parallel  \Phi   \parallel ^2_2
\end{equation}

\section{EXPERIMENTS}
\subsection{Experimental Settings}
\subsubsection{Dataset description}
We collected the most recent datasets and excluded users with fewer than five interactions to enhance the training set. The data were split into training sets and test sets, with a ratio of 70\% - 30\%, respectively. All timestamps in the datasets are recorded in their respective local time zones, ensuring temporal consistency with the actual user check-in behaviors. The statistics of the datasets are summarized in Table \ref{tab:data}. This table presents the number of users and items, the total number of interactions in the dataset, and the density, which is defined as the ratio of interactions to the product of the number of users and items.

\begin{itemize}
    \item \textbf{NYC.} The NYC dataset is widely used as a reference in research on location-based recommendation systems. It comprises user check-in records collected within New York City, including user identifiers, timestamps, and the geographical coordinates (latitude and longitude) of visited locations. 
    
    \item \textbf{Gowalla.} The Gowalla dataset serves as a widely recognized reference for location-based recommendation tasks. It contains user check-in data collected from a range of globally distributed cities, including Austin, New York, Dallas, Houston, Bangkok, Tokyo, San Antonio, and Singapore. 
    
    \item \textbf{BrightKite.} The BrightKite dataset offers comprehensive location-based user data across multiple major global regions, such as America/Chicago, Asia/Tokyo, Europe/London, and Australia/Sydney. It consists of detailed check-in records with user identifiers, timestamps, and visited locations. 
\end{itemize}

\begin{table}
	\caption{Statistics of the Experimental Datasets.}
	\label{tab:data}
	\begin{tabular}{|l|r|r|r|r|r}
		\hline
		\textbf{Dataset} & \textbf{\#User} & \textbf{\#Item} & \textbf{\#Checkin} & \textbf{Density} \\
		\hline
		NYC              & 1,083           & 9,989           & 179,468            & 1.157\%          \\
		Gowalla      & 8,685             & 13,812           & 507,197             & 0.293\%          \\
		Brightkite   & 5,766           &  14,653           & 841,511            & 0.694\%          \\
		\hline
	\end{tabular}
\end{table}

\subsubsection{Baselines} We used the same datasets and repeated experiments on the following baseline models to evaluate their performance.
\begin{itemize}
	\item \textbf{NGCF} \cite{wang2019ngcf} represents users and items as nodes in a graph and uses GCN layers to learn high-order interaction features by propagating and aggregating information across layers.
	\item \textbf{GCN-LOC} \cite{tiis:100957} enhances graph-based methods by retrieving relational attributes, improving link prediction and recommendation tasks.
	\item \textbf{LightGCN} \cite{he2020lightgcn} simplifies traditional GNNs by removing transformation matrices and nonlinear activations, focusing on effective propagation and aggregation of information.
	\item \textbf{NCL} \cite{Lin2022ImprovingGC} (Neighborhood-enriched Contrastive Learning) enriches graph filtering by leveraging structural and semantic neighbors, significantly improving the accuracy of recommendation in sparse datasets.
	\item \textbf{EEDN} \cite{Wang2023EEDN} (Enhanced Encoder-Decoder Network) integrates local and global contexts via an encoder-decoder architecture, achieving state-of-the-art performance in POI recommendations.
	\item \textbf{BIGCF} \cite{10.1145/3626772.3657738} (Bilateral Intent-guided Graph Collaborative Filtering) models individual and collective intents using a Gaussian-based graph strategy and self-supervised contrastive learning to address sparse feedback challenges.
	\item \textbf{CaDRec} \cite{10.1145/3626772.3657799} (Contextualized and Debiased Recommender Model) mitigates over-smoothing in GNNs and interaction biases using hypergraph convolutions, bias modeling, and positional encoding.
	\item \textbf{TransGNN} \cite{10.1145/3626772.3657721} combines Transformer and GNN layers with positional encoding and node sampling to aggregate global information and improve scalability and accuracy.
\end{itemize}

\subsubsection{Evaluation Metrics}
To assess the performance of our recommendation system, we employ four widely used evaluation metrics: Precision@k, Recall@k, NDCG@k, and Accuracy. These metrics offer a comprehensive assessment view of the system's effectiveness and ranking quality.

	      
	      
	      

\begin{itemize}
	\item \textbf{Precision@k}: The fraction of the top-k retrieved items that are relevant. Indicates the system’s ability to avoid irrelevant recommendations within the top results.
	\item \textbf{Recall@k}: The fraction of relevant items that are found among the top-k retrieved results. Reflects the system’s ability to capture relevant information within a limited recommendation list.
	\item \textbf{NDCG@k}: Measures ranking quality of top-$k$ results, giving more weight to relevant items appearing earlier.
	\item \textbf{Accuracy}: The proportion of correct predictions over all predictions. Provides an overall performance measure.
\end{itemize}

All results are averaged over 5 independent runs. We apply pairwise t-tests between SEP-GCN and each baseline. Values from our model that show statistically significant improvements are marked at the 99\% confidence level ($p < 0.01$).

\subsubsection{Hyperparameter Settings}
Following common practice in prior work, we adopt standard hyperparameter values without conducting extensive tuning, as our focus is not on hyperparameter optimization. Specifically, we set the learning rate to 0.001, the L2 regularization coefficient to $1 \times 10^{-5}$, and use three layers in the Light Graph Convolutional Network (LGCN), each with an embedding size of 64. We also follow the same early stopping criteria used in NGCF and LightGCN to ensure fair and consistent comparisons.

\subsection{Overall Performance and Comparative Evaluation of SEP-GCN}
\begin{table*}
	\caption{Overall performance comparison across data sets and metrics. The numbers in bold indicate statistically significant improvement (p < .01) by the pairwise t-test }
	\label{tabResult}
        \begin{adjustbox}{width=\textwidth} 
	\begin{tabular}{|c|c|c|c|c| c|c|c| c|c|c| c|c|}
		\hline
		\textbf{Dataset} & \multicolumn{4}{c|}{\textbf{NYC}} & \multicolumn{4}{c|}{\textbf{Gowalla}} & \multicolumn{4}{c|}{\textbf{Brightkite}}\\
		         & recall              & prec.               & ndcg                & acc.                & recall              & prec.               & ndcg                & acc.                & recall              & prec.               & ndcg                & acc.                \\ 
		\hline   
		\multicolumn{10}{l}{\textit{Top-5 results}}\\
		\hline
		LightGCN & 0.0116             & 0.0458             & 0.0491             & 0.1943             & 0.0364             & 0.0675             & 0.0749             & 0.2562             & 0.0107             & 0.0241             & 0.0268             & 0.1739             \\
		GCN-LOC & 0.0126             & 0.0493             & 0.0529             & 0.2123             & 0.0374             & 0.0682             & 0.0752             & 0.2611             & 0.0112             & 0.0252             & 0.0274             & 0.1811             \\
		NCL      & 0.0125             & 0.0495             & 0.0532             & 0.2073             & 0.0375             & 0.0680             & 0.0752             & 0.2651             & 0.0113             & 0.0258             & 0.0279             & 0.1805             \\
		BIGCF    & 0.0128             & 0.0521             & 0.0545             & 0.2134             & 0.0382             & 0.0695             & 0.0762             & 0.2769             & 0.0119             & 0.0273             & 0.0290             & 0.1926             \\
		EEDN     & 0.0127             & 0.0505             & 0.0535             & 0.2076             & 0.0379             & 0.0705             & 0.0754             & 0.2765             & 0.0121             & 0.0280             & 0.0293             & 0.2001             \\
		CaDRec   & 0.0128             & 0.0517            & 0.0550             & 0.2152 & \underline{0.0400}             & \underline{0.0734}             & 0.0788             & \underline{0.2900}             & 0.0128            & \underline{0.0298}             & 0.0311             & 0.2145             \\
		TransGNN & \underline{0.0130} & \underline{0.0519} & \underline{0.0559} & \underline{0.2161}             & 0.0399 & 0.0733 & \underline{0.0789} & 0.2886 & \underline{0.0128} & 0.0295 & \underline{0.0311} & \underline{0.2173} \\
		SEP-GCN  & \textbf{0.0133}    & \textbf{0.0521}    & \textbf{0.0581}    & \textbf{0.2324}    & \textbf{0.0440}    & \textbf{0.0748}    & \textbf{0.0832}    & \textbf{0.2988}    & \textbf{0.0133}    & \textbf{0.0309}    & \textbf{0.0322}    & \textbf{0.2262}    \\ \hline
		Improv.  & 2.30\%              & 0.38\%              & 3.93\%              & 7.54\%              & 10.00\%              & 1.91\%              & 5.45\%              & 3.03\%              & 3.91\%              & 3.69\%              & 3.54\%              & 4.09\%              \\ \hline

		\multicolumn{10}{l}{\textit{Top-20 results}}\\
		\hline    
		LightGCN & 0.0274             & 0.0276             & 0.0382             & 0.3958             & 0.0714             & 0.0354             & 0.0809             & 0.4563             & 0.0253             & 0.0147             & 0.0265             & 0.4088             \\
		GCN-LOC & 0.0286             & 0.0298             & 0.0407             & 0.4206             & 0.0742             & 0.0357             & 0.0817             & 0.4597             & 0.0266             & 0.0155             & 0.0277             & 0.4288             \\
		NCL      & 0.0282             & 0.0296             & 0.0404             & 0.4116             & 0.0759             & 0.0368             & 0.0814             & 0.4617             & 0.0274             & 0.0160             & 0.0283             & 0.4292             \\
		BIGCF    & 0.0295             & 0.0304            & 0.0425             & 0.4208             & 0.0796             & 0.0386             & 0.0834             & 0.4719             & 0.0293             & 0.0169             & 0.0302             & 0.4454             \\
		EEDN     & 0.0283             & 0.0304             & 0.0414             & 0.4083             & 0.0811             & 0.0393             & 0.0835             & 0.4680             & 0.0295             & 0.0173             & 0.0309             & 0.4545             \\
		CaDRec   & \underline{0.0304}             & \underline{0.0307} & 0.0426             & 0.4272             & 0.0853             & \underline{0.0413}             & 0.0874             & \underline{0.4932}             & 0.0308             & 0.0181             & 0.0328             & 0.4859             \\
		TransGNN & 0.0300 & 0.0304             & \underline{0.0427} & \underline{0.4384} & \underline{0.0880} & 0.0410 & \underline{0.0879} & 0.4903 & \underline{0.0309} & \underline{0.0188} & \underline{0.0331} & \underline{0.4950} \\
		SEP-GCN  & \textbf{0.0332}    & \textbf{0.0331}    & \textbf{0.0432}    & \textbf{0.4624}    & \textbf{0.1011}    & \textbf{0.0462}    & \textbf{0.0917}    & \textbf{0.5291}    & \textbf{0.0333}    & \textbf{0.0206}    & \textbf{0.0368}    & \textbf{0.5317}    \\ \hline
		Improv.  & 9.21\%              & 7.82\%              & 1.17\%              & 5.47\%              & 14.89\%              & 11.86\%              & 4.32\%              & 7.28\%              & 7.77\%              & 9.57\%              & 11.18\%              & 7.42\%              \\ 
		
		\hline
	\end{tabular}
	\label{figresult}
	\Description[Overall result]{}
    \end{adjustbox}
\end{table*}

The experimental results presented in Table~\ref{tabResult} clearly demonstrate the superior performance of the proposed SEP-GCN model over all baseline methods in three benchmark datasets: NYC, Gowalla, and Brightkite. SEP-GCN consistently achieves the highest scores in all four key evaluation metrics: recall, precision, NDCG, and accuracy, under different ranking thresholds (Top-5 and Top-20). The findings confirm the effectiveness of the model in both capturing user preferences and modeling spatial-temporal dependencies. The detailed analysis is summarized as follows.

\begin{itemize}
    \item \textbf{Top-5 Results:} SEP-GCN achieves significant improvements across all datasets and metrics. For example, it obtains the highest recall@5 of 0.0440 in the Gowalla data set, representing a 10.00\% improvement over the best baseline (CaDRec: 0.0400). In NYC, SEP-GCN reaches recall@5 of 0.0133, outperforming TransGNN (0.0130) and LightGCN (0.0116). Brightkite shows a similar trend, with SEP-GCN achieving 0.0133 in recall@5, again the highest among all methods. In terms of precision, SEP-GCN also delivers strong gains, notably 0.2324 in NYC (+7. 54\% over TransGNN), 0.2988 in Gowalla, and 0.2262 in Brightkite.
    
    
    \item \textbf{Top-20 Results:} SEP-GCN continues to dominate with notable margins. For example, it achieves recall@20 of 0.1011 on Gowalla, significantly outperforming LightGCN (0.0714) and TransGNN (0.0880). On Brightkite, the NDCG@20 value is 0.0368, which is a substantial improvement over EEDN (0.0309), highlighting the model's robustness in deeper rankings. The accuracy@20 metric also improves consistently, reaching up to 0.5291 on Gowalla and 0.5317 on Brightkite.

    
\end{itemize}

In conclusion, SEP-GCN is a robust and highly effective model for location-based recommendation. It consistently outperforms state-of-the-art baselines such as LightGCN, BIGCF, and TransGNN, and exhibits strong generalizability across diverse data conditions. Its superior performance in all metrics and ranking thresholds positions it as a leading solution for real-world POI recommendation systems.

\subsection{In-depth studies of SEP-GCN}
We conduct further experiments to analyze the impact of various components of the model architecture, including spatial ablation and an analysis of the number of layers. Furthermore, we evaluated the performance of SEP-GCN in sparse datasets to demonstrate its effectiveness in handling sparsity issues.

\subsubsection{Scalability to Large-Scale Datasets}

\begin{table}[ht]
\centering
\caption{Comparison between Regional and Full Datasets}
\label{tab:dataset_comparison}
\begin{adjustbox}{width=\columnwidth}
\begin{tabular}{l|r|r|r|r}
\toprule
\textbf{Dataset} &  \textbf{\#User} & \textbf{\#Item} & \textbf{\#Checkin} & \textbf{Density} \\
\midrule
Gowalla@BKK & 981   & 2,544  & 34,627   & 0.953\% \\
Gowalla Full     & 8,685 & 13,812 & 507,197  & 0.293\% \\
Brightkite@TYO & 1,308 & 5,835  & 124,546  & 1.134\% \\
Brightkite Full     & 5,766 & 14,653 & 841,511  & 0.694\% \\
\bottomrule
\end{tabular}
\end{adjustbox}
\end{table}

To improve the robustness of our research, we extended our experiments from regional subsets (Gowalla at Bangkok and Brightkite at Tokyo) to full datasets. Although regional data allowed fine-grained analysis, they limited generalizability. We restructured our framework to support cloud-based GPU training, allowing us to scale to the full Gowalla and Brightkite datasets with 507k and 841k check-ins, respectively - significantly larger than their regional counterparts, as shown in Table \ref{tab:dataset_comparison}. In both cases, SEP-GCN consistently outperforms the TransGNN baseline in all metrics: recall, precision, NDCG, and accuracy.

\begin{table}[ht]
\centering
\caption{Comparison of SEP-GCN performance on Regional vs. Full Datasets (Top@20)}
\label{tab:sep_gcn_subset}
\begin{adjustbox}{width=\columnwidth}
\begin{tabular}{l|l|c|c|c|c}
\toprule
\textbf{Dataset} & \textbf{Model} & \textbf{Recall} & \textbf{Precision} & \textbf{NDCG} & \textbf{Accuracy} \\
\midrule
\multirow{3}{*}{Gowalla@BKK} 
  & TransGNN & 0.1015 & 0.0275 & 0.0863 & 0.4182 \\
  & SEP-GCN                   & 0.1071 & 0.0298 & 0.0915 & 0.4322 \\
  & \textit{Improvement}   & 5.51\% & 8.36\% & 6.02\% & 3.35\%  \\
 \midrule
\multirow{3}{*}{Gowalla Full} 
  & TransGNN & 0.0880 & 0.0410 & 0.0879 & 0.4903 \\
  & SEP-GCN                 & 0.1011 & 0.0462 & 0.0917 & 0.5291 \\
  & \textit{Improvement} & 14.89\% & 12.68\% & 4.32\% & 7.91\% \\

 \midrule
\multirow{3}{*}{Brightkite@TYO} 
  & TransGNN & 0.0691 & 0.0179 & 0.0504 & 0.2453\\
  & SEP-GCN                   & 0.0741 & 0.0185 & 0.0529 & 0.2515 \\
  & \textit{Improvement}   & 7.23\% & 3.35\% & 4.96\% & 2.53\% \\
  \midrule
\multirow{3}{*}{Brightkite Full} 
  & TransGNN & 0.0309 & 0.0188 & 0.0331 & 0.4950  \\
  & SEP-GCN                   & 0.0333 & 0.0206 & 0.0368 & 0.5317 \\
  & \textit{Improvement}   & 7.77\% & 9.57\% & 11.18\% & 7.42\% \\
\bottomrule
\end{tabular}
\end{adjustbox}
\end{table}

On local datasets such as Gowalla@BKK and Brightkite@TYO, SEP-GCN shows clear advantages, with moderate improvements (e.g. +5. 51\% recall in Gowalla@BKK and +7. 23\% in Brightkite@TYO), demonstrating its effectiveness in capturing fine-grained spatial-temporal patterns within concentrated geographical regions.

When scaling to the full datasets, SEP-GCN not only retains but even surpasses its previous performance, achieving up to +14.89\% recall and +11.18\% NDCG improvements, as shown in Table \ref{tab:sep_gcn_subset}. This enhancement is particularly noteworthy given the significantly higher sparsity of the full datasets (e.g., 0.293\% for Gowalla Full vs. 0.953\% for Gowalla@BKK).

Interestingly, the model performs even better on the full datasets than on the regional ones. This is attributed to the optimized design of SEP-GCN, which considers both geographical distance and local time zones, a crucial advantage when the data span multiple cities and time zones. These results highlight the scalability and robustness of SEP-GCN in modeling complex, sparse user-item interactions across diverse spatial settings, making it a practical solution for location-sensitive real-world recommendation systems.

These results demonstrate the scalability and robustness of SEP-GCN to real-world data distributions. In addition, the consistent improvement from region-specific to complete datasets validates the generalizability of the model and the practical readiness to deploy it. The positive effect of this data set transition reinforces architectural design choices in SEP-GCN and confirms its effectiveness under diverse geographical and behavioral conditions.

\subsubsection{Contribution of Spatial and Temporal in SEP}
We conduct experiments using SEP-GCN and chose LightGCN as a representative GCN-based model due to its architectural simplicity, which makes it suitable for isolating the effects of spatial and temporal components. For LightGCN with spatial information, we integrate a location-aware weight matrix into the adjacency matrix $A = \begin{bmatrix} 0 & R \\ R^\top & L \end{bmatrix}$, where $L$ encodes spatial distances. In the Temporal-only variant of SEP-GCN, we omit the construction of the geographical distance-based matrix $\mathcal{X}$. Conversely, in the Spatial-only variant, we remove the time slot intersection condition from Definition 5.

The performance comparison of different variants is presented in Table \ref{tab:ablation1}. Adding spatial distance to LightGCN yields minor improvements on NYC but degrades performance on Gowalla and Brightkite, suggesting that spatial signals alone may introduce noise. In contrast, removing distance from SEP-GCN causes a significant performance drop, demonstrating its effective use of spatial information—not as a static input, but as a filter over temporal relations.

Temporal clustering also plays a vital role. Without it, SEP-GCN cannot adapt the similarity matrix $\mathcal{X}$ during training, leading to reduced performance. These results confirm that the joint integration of spatial and temporal signals, using distance as a contextual filter, is key to the effectiveness of SEP-GCN.

\begin{table}[ht]
	\centering
	\caption{Performance comparison of SEP-GCN variants.}
	\label{tab:ablation1}
	\begin{adjustbox}{width=\columnwidth} 
		\begin{tabular}{|c|c|cc|cc|cc|}
			\hline
			\multirow{2}{*}{Model} & \multirow{2}{*}{Variants} & \multicolumn{2}{c|}{NYC} & \multicolumn{2}{c|}{Gowalla} & \multicolumn{2}{c|}{Brightkite} \\ \cline{3-8} 
			  &              & R@20             & N@20             & R@20             & N@20             & R@20             & N@20             \\ \hline
			\multirow{2}{*}{LightGCN} 
			  & w/ Spatial   & 0.0278          & 0.0390          & 0.0709          & 0.0805          & 0.0250          & 0.0262          \\ 
			
			  & Original     & 0.0274          & 0.0382          & 0.0714          & 0.0809          & 0.0253          & 0.0265          \\ 
			\hline
			\multirow{3}{*}{SEP-GCN} 
			  & only Temporal  & 0.0253          & 0.0308          & 0.0674          & 0.0681          & 0.0213          & 0.0248          \\ 
			    
			  & only Spatial & 0.0306          & 0.0361          & 0.0837          & 0.0729          & 0.0288          & 0.0279          \\ 
			
			  & Original     & \textbf{0.0332} & \textbf{0.0432} & \textbf{0.1011} & \textbf{0.0917} & \textbf{0.0333} & \textbf{0.0368} \\ 
			\hline
		\end{tabular}
	\end{adjustbox}
	\label{tab:transgnn_variants}
\end{table}

\subsubsection{The effectiveness on number of layers}
\begin{figure}[ht]
    \centering
    \includegraphics[width=0.5\textwidth]{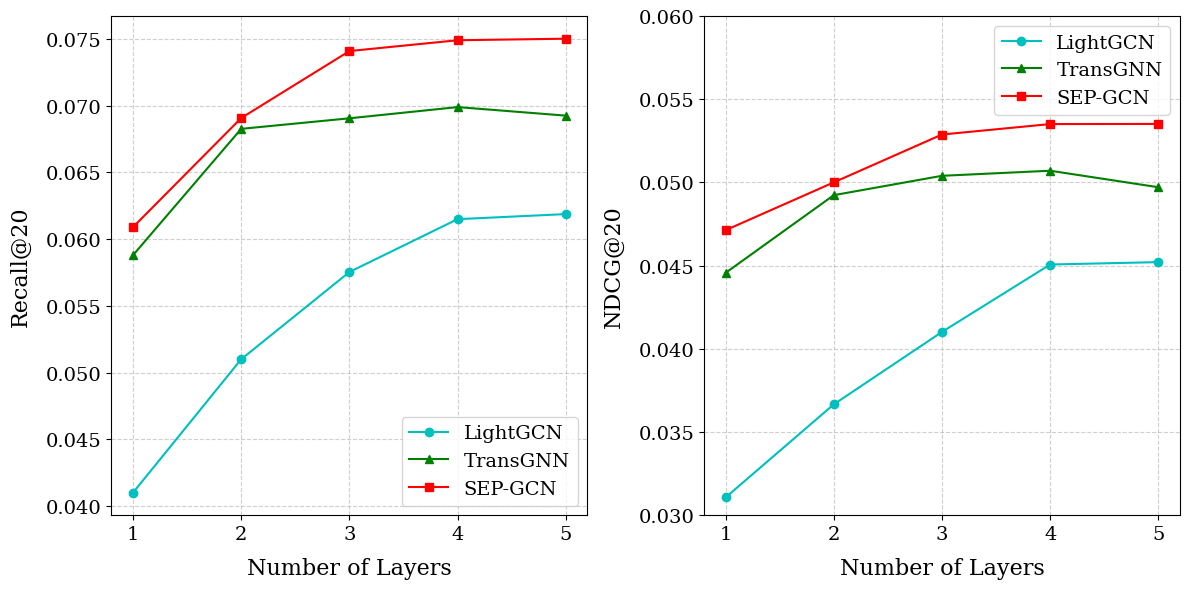}
    \caption{Comparison of Recall@20 and NDCG@20 for different models and number of layers.}
    \label{fig:comparison_graph}
    \Description[Figure compares the performance of LightGCN, TransGNN, and SEP-GCN in terms of Recall@20 and NDCG@20 across varying numbers of layers.]{Figure presents a comparison of three models—LightGCN, TransGNN, and SEP-GCN—on two key metrics: Recall@20 (left) and NDCG@20 (right), as the number of graph convolutional layers increases from 1 to 5. SEP-GCN demonstrates superior performance across both metrics, achieving the highest scores at most layer depths. TransGNN shows competitive performance but falls short of SEP-GCN, particularly in deeper layers. LightGCN, while efficient, trails behind both models, with a noticeable gap in performance for both Recall@20 and NDCG@20. This highlights the advantages of SEP-GCN's edge-enhanced mechanisms in effectively capturing user-item relationships.}
\end{figure}
We compare SEP-GCN with LightGCN and TransGNN to evaluate the effect of propagation layers on the Gowalla dataset. As shown in Figure~\ref{fig:comparison_graph}, SEP-GCN consistently outperforms other models from the first layer by effectively capturing high-quality signals without relying on deep propagation. This advantage comes from the early integration of temporal and structural information, which also helps mitigate the over-smoothing issue.

Traditional GCN-based models such as LightGCN typically improve performance up to 2--3 layers but then degrade as the number of layers increases. TransGNN, while initially competitive, also suffers from over-smoothing at deeper layers. In contrast, SEP-GCN maintains stable and high performance even beyond three layers, demonstrating its robustness and ability to capture deeper graph-level features. Figure~\ref{fig:comparison_graph} illustrates this trend, highlighting the performance drop in traditional models compared to the steady behavior of SEP-GCN.

\subsubsection{The effectiveness on spare dataset}

To evaluate the performance of SEP-GCN under sparse data conditions, we constructed two variants of each dataset using $k$-core filtering with thresholds of 5 and 10. The 5-core datasets exhibit extreme sparsity: 98.843\% for NYC, 99.883\% for Gowalla and 99.501\% for Brightkite, providing a challenging setting for recommendation models.

The experimental results show that SEP-GCN performs robustly under sparse data conditions, achieving improvements ranging from 5.6\% to 23.2\% in the 5-core scenario. The detailed performance comparison is shown in Table \ref{tab:ablation2}. These gains are substantially larger than those observed on the 10-core datasets, where improvements range from only 1.17\% to 14.89\%. This shows its strong ability to capture user preferences even with limited interactions, addressing near-cold-start challenges where strong existing methods such as TransGNN fail. By integrating spatial and temporal information, the SEP graph enhances connectivity in the interaction space, effectively linking sparse or isolated nodes that conventional interaction-only methods cannot reach.


\begin{table}[ht]
	\centering
	\caption{Performance comparison on sparsity of dataset}
	\label{tab:ablation2}
	\begin{adjustbox}{width=\columnwidth} 
		\begin{tabular}{|c|c|cc|cc|}
			\hline
			\multirow{2}{*}{Datasets} & \multirow{2}{*}{Variants} & \multicolumn{2}{c|}{5-core} & \multicolumn{2}{c|}{10-core} \\ \cline{3-6} 
			  &          & R@20             & N@20             & R@20             & N@20             \\ \hline
			\multirow{3}{*}{NYC} 
                & TransGNN & 0.0331          & 0.0402          & 0.0300          & 0.0427          \\ 
			  & SEP-GCN  & \textbf{0.0392} & \textbf{0.0425} & \textbf{0.0332} & \textbf{0.0432} \\ 
			  & Improv.  & 18.3\%           & 5.6\%           & 10.67\%            & 1.17\%            \\ \hline
			\multirow{3}{*}{Gowalla} 
                & TransGNN & 0.0915          & 0.0784          & 0.0880          & 0.0879          \\ 
			  & SEP-GCN  & \textbf{0.1128} & \textbf{0.0923} & \textbf{0.1011} & \textbf{0.0917} \\ 
			  & Improv.  & 23.2\%           & 17.7\%           & 14.89\%           & 4.32\%            \\ \hline
			\multirow{3}{*}{Brightkite} 
                & TransGNN & 0.0316         & 0.0272          & 0.0309          & 0.0331          \\ 
			  & SEP-GCN  & \textbf{0.0374} & \textbf{0.0324} & \textbf{0.0333} & \textbf{0.0368} \\ 
			  & Improv.  & 18.4\%           & 19.1\%           & 10.05\%           & 11.18\%           \\ \hline
		\end{tabular}
	\end{adjustbox}
	\label{tab:k-core}
\end{table}

\section{CONCLUSION}
In this work, we proposed \textbf{SEP-GCN}, a novel framework that enhances traditional Graph Convolutional Networks by incorporating both temporal and geographical associations to improve recommendation accuracy. By systematically analyzing item interactions within shared time windows and spatial proximities, our method constructs context-enriched relational representations that more effectively capture user preferences.

The integration of these contextual signals through an edge-enhanced graph convolutional mechanism enables more nuanced and robust predictions. Extensive experiments on benchmark datasets demonstrate that SEP-GCN consistently outperforms state-of-the-art baselines, showcasing its predictive strength and adaptability across diverse recommendation scenarios.

These results underscore the importance of modeling contextual similarity in graph-based recommendation systems to address the challenges of information overload and dynamic user behavior. SEP-GCN lays a strong foundation for future research on context-aware recommendation, paving the way for more personalized, scalable, and user-centric systems.

\printbibliography

\end{document}